\documentclass[prl,amsmath,amssymb,twocolumn,showpacs,aps]{revtex4}
\usepackage{graphicx}
\usepackage{bm}

\begin{document}

\title{Magneto-oscillations due to electron-electron interactions in
the ac conductivity of a 2D electron gas}

\author{T. A. Sedrakyan and M. E. Raikh}

\affiliation{ Department of Physics, University of Utah, Salt Lake
City, UT 84112}

\begin{abstract}

Electron-electron interactions give rise to the correction,
$\delta\sigma^{int}(\omega),$ to the ac  magnetoconductivity,
$\sigma(\omega)$,  of a clean 2D electron gas that is periodic in
$\omega_c^{-1}$, where $\omega_c$ is the cyclotron frequency.
Unlike conventional harmonics of the cyclotron resonance,
which are periodic with $\omega$, this correction is periodic with
$\omega^{3/2}$. Oscillations in
$\delta\sigma^{int}(\omega)$
develop at low magnetic fields, $\omega_c\ll\omega$,
when the conventional harmonics are suppressed by the disorder.
Their origin is
a {\em double} backscattering of an electron from the
impurity-induced Friedel oscillations. During the time
$\sim\omega^{-1}$ between the two backscattering events the
electron travels only a {\em small portion} of the Larmour circle.

\end{abstract}


\maketitle

{\noindent \it Introduction.} Originally, the  cyclotron resonance
(and its harmonics) in the ac conductivity, $\sigma(\omega)$, of
the 2D electron gas had been detected by measuring the
transmission of the microwave radiation \cite{koch74}. In the
recent experiment on high-mobility samples \cite{zudov01}, it was
demonstrated that this resonance, together with harmonics, also
manifests itself in the dc magnetoresistance under microwave
illumination, i.e., in the photoconductivity. A spectacular
strength of this effect,
and, in particular, observation of the zero-resistance state,
above a certain intensity of illumination
\cite{zudov03,mani02,dorozhkin03}, had attracted a steady interest
of the researchers to the ac-response of a high-mobility electron
gas in a weak magnetic field, $B$. Unlike the conventional
Shubnikov-de Haas oscillations of the dc magnetoresistance,  which
vanish with temperature as $\exp(-2\pi^2T/\omega_c)$, where
$\omega_c$ is the cyclotron quantum, the magneto-oscillations of
$\sigma(\omega)$ survive at high temperature
\cite{mirlin03,mirlin04}. The shape of these oscillations is given
by \cite{mirlin03,mirlin04}
\begin{equation}
\label{standard}
\frac{\delta\sigma_{\pm}(\omega)}{\sigma_{\pm}(\omega)}=2\cos\Biggl(\frac{2\pi\omega}{\omega_c}\Biggr)
\exp\left(-\frac{2\pi}{\omega_c\tau}\right),
\end{equation}
where $\tau$ is the scattering time, and $\sigma_{\pm}(\omega)$ is
related to the dc conductivity $\sigma_0$ as
$\sigma_{\pm}(\omega)={\sigma_0}/{\left[2+2(\omega\pm\omega_c)^2\tau^2\right]}$.
Classically, the meaning of the damping factor
$\delta^2=\exp\bigl(-2\pi/\omega_c\tau\bigr)<1$ is the probability
for an electron to execute the entire Larmour circle, $2\pi
R_{\mbox{\tiny L}}$, without being scattered.
Oscillations Eq.~(\ref{standard})
is a {\em single-electron} effect. In
converting of these oscillations into the oscillating dc
photoconductivity \cite{mirlin04,vavilov04,mirlin05,long},  the
electron-electron interactions enter as a source of relaxation of
the oscillatory part of the distribution function.

In the present paper we demonstrate that
interactions {\em by themselves} give rise to
the  oscillatory contribution, $\delta\sigma^{int}(\omega)$,
to the {\em linear} ac conductivity,
$\sigma(\omega)$,
at frequencies much higher than in
Eq.~(\ref{standard}).
To contrast this
contribution to Eq.~(\ref{standard}), we present
$\delta\sigma^{int}(\omega)$ in the form

\begin{figure}[t]
\centerline{\includegraphics[width=45mm,angle=0,clip]{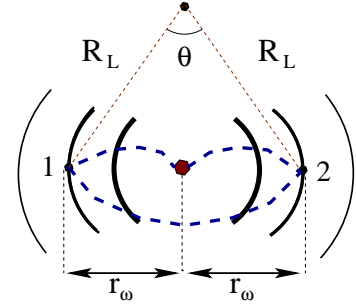}}
\caption{Illustration of the $B$-dependent contribution,
$\delta\Theta_{\mbox{\tiny B}}(r)$, to the phase of the
polarization operator; $R_{\mbox{\tiny
L}}[\theta-2\sin(\theta/2)]$ is the elongation of the
semiclassical trajectory due to the field-induced curving. The
origin of oscillating magnetoconductivity is the scattering from
the Friedel oscillations (arcs of decreasing thickness), at points
1 and 2, located {\em symmetrically} with respect to the impurity
shown with a 
big dot.}
\end{figure}

\begin{eqnarray}
\label{predict}
\!\frac{\delta\sigma^{int}(\omega)}{\sigma(\omega)}\propto
\cos\Biggl(\!\!\!\!&C&\!\!\!\!_{\omega}\frac{\omega}{\omega_c}-\frac{\pi}{4}\Biggr)
\exp\left[-\frac{3C_{\omega}}{\omega_c}\left(\frac{1}{\tau}+2\pi
T\right)\right],\nonumber\\
\end{eqnarray}
where
\begin{equation}
\label{comega}  C_{\omega}=\left[\frac{32\omega}{27E_{\mbox{\tiny
F}}}\right]^{1/2},
\end{equation}
 and $E_{\mbox{\tiny F}}$ is
the Fermi energy. Since $C_{\omega}$ is small, the correction
Eq.~(\ref{predict}) develops oscillations at much smaller magnetic
fields $\omega_c \sim C_{\omega}\omega\ll \omega$ than
Eq.~(\ref{standard}). At such fields, the damping factor in
Eq.~(\ref{standard}) is $\sim \exp\left(-2\pi/C_{\omega}\right)$,
i.e., the conventional oscillations are completely washed out.

{\noindent \it Qualitative picture.} The origin of the
oscillations Eq.~(\ref{predict}) lies in a peculiar modification
by the interactions  of the impurity scattering
in a weak magnetic field. Conventionally,
\cite{Dolgopolov,Narozhny,gornyi04} this modification amounts to
the additional scattering \cite{rudin} from the Friedel
oscillations of the electron density, created by the impurity.
Such a modification {\em does not} lead to the anomalous
sensitivity to low $B$. However, as we demonstrate below, this
sensitivity emerges in the {\em second order} in the
electron-electron interaction strength. The corresponding
second-order processes are illustrated in Figs.~1 and 2. They are:
\noindent({\em i}). Photoexcited electron emits a virtual pair,
which is subsequently annihilated. Impurity scatters {\em not} the
original electron, but rather the impurity scattering occurs
between the states, {\em constituting the pair}, prior to
annihilation. Diagram $b$ in Fig.~2 describes this process.
\noindent({\em ii}) Electron is {\em not} scattered {\em directly}
by the impurity, but rather experiences a {\em double}
backscattering from the impurity-induced Friedel oscillations, as
illustrated in Fig.~1, and also by the diagram $a$ in Fig.~2.

\begin{figure}[t]
\centerline{\includegraphics[width=75mm,angle=0,clip]{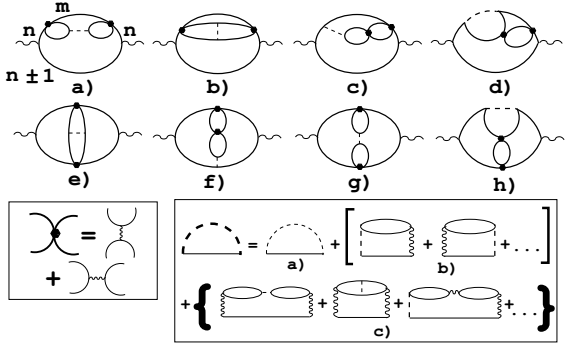}}
\caption{Diagrams contributing to the second-order interaction
correction to the ac magnetoconductivity. Dashed lines denote the
impurity scattering. Dots in the vertices combine two types of the
interaction matrix elements, as shown in the left inset. Diagrams
b and e describe the impurity scattering of the {\em secondary}
electron (hole); other diagrams describe double scattering of {\em
photoexcited} electron (hole) by the Friedel oscillation. Right
inset: diagram (a) for purely disorder-induced part of the
self-energy, $\Sigma_n$, is plotted together with representative
diagrams (b) and (c) for the interaction-induced self-energy,
$\delta\Sigma_n^{int}$. 
Only diagrams (c), describing {\em
two} electron-electron scattering processes, contribute to the
interaction-induced magneto-oscillations.}
\end{figure}

We demonstrate that the corrections to the conductivity from both
these processes {\em oscillate} with magnetic field according to
Eq.~(\ref{predict}). The oscillations reflect the fact that, for
both processes, the dominant contribution to the double
backscattering cross-section comes from two ``distinguished''
points that are located {\em symmetrically} with respect to the
impurity  at {\em certain} well-defined distance, $r_{\omega}\sim
C_{\omega}R_{\mbox{\tiny L}}$, see Fig.~1. Then the argument of
the cosine in Eq.~(\ref{predict}) can be interpreted as a product
$\omega t_{\omega}$, where $t_{\omega}=r_{\omega}/v_{\mbox{\tiny
F}}$ is the time, during which the electron with Fermi velocity,
$v_{\mbox{\tiny F}}$, travels the distance $r_{\omega}$.

{\noindent \it  Derivation.} Although the interaction-induced
oscillations  come from small distances, $r_{\omega} \ll
R_{\mbox{\tiny L}}$,
we nevertheless  will evaluate $\sigma(\omega)$ in the Landau
gauge
to demonstrate how both oscillations Eq.~(\ref{standard}) and
Eq.~(\ref{predict}) emerge from the same calculation. Within the
self-consistent Born approximation,
averaging in the general expression for the diagonal conductivity

\begin{eqnarray}
\label{kubo} \sigma(\omega)=\frac{-e^2}{4\pi ^3 \Omega}
\int\limits_{-\infty}^{\infty} \!\!\frac{d\epsilon}{\omega}
\left(f_{\epsilon}-f_{\epsilon+\omega}\right)\text{Tr}{\overline{{\hat
v_x}\text{Im}{\hat G}_{\epsilon+\omega} {\hat v_x}\text{Im}{\hat
G}_{\epsilon}}}
\end{eqnarray}
is decoupled into two
averaged Green functions


\begin{eqnarray}
\label{GLL}
G_{\epsilon}=\sum_{n}G_n(\epsilon)=\sum_{n}\frac{1}{\epsilon-\epsilon_n-\Sigma_n(\epsilon)},
\end{eqnarray}
where $\epsilon_n=(n+\frac{1}{2})\hbar\omega_c$ are the Landau
levels, and $\Sigma_n(\epsilon)$ is the self-energy. In
Eq.~(\ref{kubo}), the bar denotes the disorder averaging, $\Omega$
is the normalization area, and $f_{\epsilon}$ is the Fermi
distribution. Upon decoupling, Eq.~(\ref{kubo}) takes a familiar
form
\begin{eqnarray}
\label{sigma} \sigma_{\pm}(\omega)=\frac{e^2v_{\mbox{\tiny
F}}^2\omega_c\nu_0}{8\pi}\int\!\frac{d\epsilon}{\omega}
\Bigl(f_{\epsilon}-f_{\epsilon+\omega}\Bigr)\nonumber\\
\times \sum_n\text{Im}G_{n\pm 1}(\epsilon+\omega)\;
\text{Im}G_{n}(\epsilon),\qquad\qquad\qquad
\end{eqnarray}
where $\nu_0$ is the 2D density of states. For high Landau levels,
$\left(E_{\mbox{\tiny F}}/\hbar\omega_c\right) \gg 1$, in the
first approximation in $\delta < 1$, the self-energy can be
replaced by its zero-field value, $i/2\tau$. Then
Eq.~(\ref{sigma}) readily reproduces
the Drude conductivity. In order to capture the oscillatory ac
magnetoconductivity, in the next approximation, one should take
into account the ``quantum'' correction,
$\delta\Sigma_n^Q(\epsilon)\propto \delta \exp\left(-2\pi
i\epsilon/\omega_c\right)$, to the self-energy due to the
discreteness of the Landau levels, as well as the interaction
correction, $\delta\Sigma_n^{int}(\epsilon)$. Since both
corrections are smaller than $1/\tau$, they cause a small
correction to the Green functions Eq.~(\ref{GLL}) of the form
\begin{eqnarray}
\label{correction} \delta
G_n(\epsilon)=\frac{\delta\Sigma_n^Q(\epsilon)+\delta\Sigma_n^{int}}
{\left(\epsilon-\epsilon_n-\frac{i}{2\tau}\right)^2}.
\end{eqnarray}
The first and the second terms in Eq.~(\ref{correction}) give rise
to the oscillations Eq.~(\ref{standard}) and Eq.~(\ref{predict}),
respectively. However, to reproduce these oscillations the
``quantum'' and the interaction corrections should be handled
differently.
To reproduce
Eq.~(\ref{standard}), upon substituting Eq.~(\ref{correction})
into Eq.~(\ref{sigma}), one should keep the product,
$\delta\Sigma_{n+1}^Q(\epsilon +\omega)
\bigl[\delta\Sigma_{n}^Q(\epsilon)\bigr]^{\ast}$.
It contains the oscillating term $\propto \exp\left(-2\pi i
\omega/\omega_c\right)$, which {\em does not} depend neither on
$n$ nor on $\epsilon$.
For this reason, the resulting oscillations of magnetoconductivity
are $T$-independent.
By contrast, to capture  the interaction-induced oscillations, it
is sufficient to keep $\delta\Sigma_n^{int}$ only in one of the
Green functions in Eq.~(\ref{sigma}), and its $n$-dependence is
crucial.
We will perform further calculation for
$\delta\Sigma_n^{int}(\epsilon)$ given by the first diagram of
type $c$ in Fig.~2 (inset). This is because the diagrams of type
$b$ do not cause magneto-oscillations, while the contributions of
other diagrams of type $c$ are comparable to that of the first
one, and will be addressed later.

The first diagram of type $c$ can be presented as
\begin{eqnarray}
\label{specific} \delta\Sigma_n^{int}(\epsilon)=\sum_m\frac{\vert
R_{nm}\vert^2}{\epsilon-\epsilon_m+\frac{i}{2\tau}},
\end{eqnarray}
so that the $n$-dependence is encoded in the ``matrix elements'',
$R_{nm}$. Substituting Eq.~(\ref{specific}) into
Eq.~(\ref{correction}), and then  Eq.~(\ref{correction}) into
Eq.~(\ref{sigma}) yields
\begin{eqnarray}
\label{interact}
\!\!\!\!\!\!\!\!\!\!&\delta&\!\!\!\!\!\!\sigma^{int}(\omega)\propto\int\frac{d\epsilon}{\omega}\bigl(f_{\epsilon}-f_{\epsilon+\omega}\bigr)\times
\!\!\!\!\\
&\text{Im}&\!\!\sum_{n,m}\frac{\vert
R_{nm}\vert^2}{\left(\epsilon-\epsilon_n+\frac{i}{2\tau}\right)^2
\left(\epsilon+\omega-\epsilon_{n\pm 1}-\frac{i}{2\tau}\right)
\left(\epsilon-\epsilon_m+\frac{i}{2\tau}\right)}.\nonumber
\end{eqnarray}
As a next step, we express the matrix element, $R_{nm}$, as an
integral in the coordinate space, following Fig.~2c,
$R_{nm}\propto \int d{\bf r}\psi_n^{\ast}({\bf r})\psi_m({\bf
r})\Pi_{0}({\bf r},0)$,
where $\Pi_{0}({\bf r},0)$ is  the {\em static} polarization
operator between the point ${\bf r}=0$, where the impurity is
located, and the point ${\bf r}$, where the backscattering takes
place. Our prime observation is that with such $R_{nm}$ the
relevant term in Eq.~(\ref{interact}), which has the form
\begin{eqnarray}
\label{polar}
&-&i\int d\epsilon\left(f_{\epsilon+\omega}-f_{\epsilon}\right)\\
&\times& \!\!\!\!\frac{1}{\omega-\frac{i}{\tau}}
\sum_{n,m}\frac{\psi_n^{\ast}({\bf r}_1)\psi_n({\bf
r}_2)\psi_m^{\ast}({\bf r}_2)\psi_m({\bf r}_1)}
{\left(\epsilon+\omega-\epsilon_{n\pm
1}-\frac{i}{2\tau}\right)\left(\epsilon-\epsilon_m+\frac{i}{2\tau}\right)},
\nonumber
\end{eqnarray}
again {\em reduces to the polarization operator},
$\Pi_{\omega\mp\omega_c}({\bf r}_1,{\bf r}_2)$.

In the final expression for the interaction correction we make use
of the fact that the $B$-dependence of this correction develops in
the low-field limit $\omega_c\sim C_{\omega}\omega \ll \omega$,
and replace $\omega\pm \omega_c$ by $\omega$ \cite{footnote}. We
then obtain
\begin{eqnarray}
\label{final} \frac{\delta\sigma^{int}}{\sigma(\omega)}=
\frac{\lambda^2}{\pi\nu_0^4\omega}
\int\!\! d{\bf r}_1\!\int\!\! d{\bf r}_2\;\text{Im}\Pi_{\omega}({\bf r}_1,{\bf r}_2)\nonumber\\
\times\text{Re}\;\Bigl\{\Pi_0(0, {\bf r}_1)\Pi_0({\bf
r}_2,0)\Bigr\},
\end{eqnarray}
where $\lambda$ stands for dimensionless strength of interaction,
which we assumed to be short-ranged. The numerical factor in
Eq.~(\ref{final}) will be established when all the diagrams
contributing to $\delta\sigma^{int}$ are considered (see below).

{\noindent \it Interpretation.} The form of Eq.~(\ref{final}) can
be interpreted as follows. The factor, $\text{Im}\Pi_{\omega}({\bf
r}_1,{\bf r}_2)$,  in the integrand is the density-density
response, the same as in calculation of the Drude ac
conductivity.
The second factor,
$\text{Re}\;\Bigl\{\Pi_0(0, {\bf r}_1)\Pi_0({\bf r}_2,0)\Bigr\}$,
plays the role of the spatial correlator of the effective  random
potential. By lifting the momentum conservation, this potential
enables the absorption of the ac field.
If the correlator was $\propto \delta({\bf r}_1-{\bf r}_2)$, then
the rhs of Eq.~(\ref{final}) would yield an $\omega$-independent
constant. Important is that the effective potential in
Eq.~(\ref{final}) originates from the modulation of the electron
density by the impurity, and thus {\em oscillates rapidly} with
distance. It is these Friedel oscillations that in
magnetic field
lead to  the oscillating correction, Eq.~(\ref{predict}).

{\noindent \it Oscillations.} The long-distance, $k_{\mbox{\tiny
F}}r \gg 1$, behavior of the polarization operator in coordinate
space is the following
\begin{eqnarray}
\label{operator}
\Pi_{\omega}(r)=-\frac{\pi\nu_0^2\hbar^4}{2k_{\mbox{\tiny
F}}r}\Biggl[i\vert\omega\vert+v_{\mbox{\tiny
F}}\frac{\sin\bigl\{\Theta(r)\bigr\}}{r}A\Biggl(\frac{2\pi
rT}{v_{\mbox{\tiny F}}}\Biggr)\Biggr]\nonumber\\
\times\exp\Biggl\{\frac{i\vert\omega\vert r}{v_{\mbox{\tiny
F}}}-\frac{r}{v_{\mbox{\tiny F}}\tau}\Biggr\},\qquad
\end{eqnarray}
where the function $A(x)=x/\sinh(x)$ describes the temperature
damping. In the momentum space, two contributions to
Eq.~(\ref{operator}) originate from small momentum transfer and
momentum transfer close to $2k_{\mbox{\tiny F}}$, respectively
\cite{chubukov03}. At distances $r\ll R_{\mbox{\tiny L}}$, a
nonquantizing magnetic field enters into Eq.~(\ref{operator})
through the semiclassical phase, $\Theta(r)$. This phase is
accumulated by the electron upon propagation from the point $0$ to
the point ${\bf r}$ {\em and back}. In a zero magnetic field, we
obviously have, $\Theta(r)=2k_{\mbox{\tiny F}}r$. At distances $r
\ll R_{\mbox{\tiny L}}$, the field-dependent correction \cite{we}
to $\Theta(r)$ is equal to
\begin{equation}
\label{delta} \delta\Theta_{\mbox{\tiny B}}(r)=2k_{\mbox{\tiny
F}}\delta {\cal L}-\frac{{\cal A}B}{\Phi_0}= -\frac{E_{\mbox{\tiny
F}}\omega_c^2r^3}{6v_{\mbox{\tiny F}}^3}.
\end{equation}
The origin of the correction Eq.~(\ref{delta}) is illustrated in
Fig.~1. It comes from elongation, $\delta {\cal L}=R_{\mbox{\tiny
L}}[\theta-2\sin(\theta/2)]$, of the classical electron trajectory
in magnetic field,  as well as from the Aharonov-Bohm flux into
the {\em loop} with area ${\cal A}=(\theta-\sin
\theta)R_{\mbox{\tiny L}}^2/2$. The correction Eq.~(\ref{delta})
is {\em negative}, since the Aharonov-Bohm contribution {\em
exceeds twice} the orbital contribution. We emphasize, that the
conventional way~\cite{gorkov59} of incorporating magnetic field
into the
Green's function neglects the curvature of the electron
trajectories, {\em i.e.,}
$\delta\Theta_{\mbox{\tiny B}}(r)=0$. Thus, within the approach of
Ref.~\onlinecite{gorkov59}, the oscillations Eq.~(\ref{predict})
{\em would not} emerge.

Further calculation is straightforward. Substituting
Eq.~(\ref{operator}) into Eq.~(\ref{final}), performing the
angular integration, and combining rapidly oscillating terms in
the product of three polarization operators into a ``slow'' term,
we find that  the interaction correction Eq.~(\ref{final}) can be
presented as
$\delta\sigma^{int}/\sigma(\omega)=(\lambda^2E_{\mbox{\tiny
F}}/\omega) \text{\large F}_{\tau,T}$, where the dimensionless
function $\text{\large F}_{\tau,T}(\omega,\omega_c)$ is defined as
follows

\begin{eqnarray}
\label{function1} \text{\Large F}_{\tau,T}
=\frac{1}{(\pi k_{\mbox{\tiny
F}})^{5/2}}\int\limits_0^{\infty}\!\frac{dr_1dr_2}{\bigl[r_1r_2(r_1+r_2)\bigr]^{3/2}}
A\left(\frac{2\pi r_1T}{v_{\mbox{\tiny
F}}}\right)\times\;\; \nonumber\\
\!\!\!\!\!\!\!A\left(\frac{2\pi r_2T}{v_{\mbox{\tiny
F}}}\right)A\left(\frac{2\pi (r_1+r_2)T}{v_{\mbox{\tiny
F}}}\right) \exp\left\{-\frac{2(r_1+r_2)}{v_{\mbox{\tiny
F}}\tau}\right\}\times\;\;\\
\cos\Biggl[ \frac{\omega (r_1\!+\!r_2)}{v_{\mbox{\tiny
F}}}\!+\!\delta\Theta_{\mbox{\tiny B}}(r_1\!+\!r_2)\!-\!
\delta\Theta_{\mbox{\tiny B}}(r_1)\!-\!\delta\Theta_{\mbox{\tiny
B}}(r_2)\!-\!\frac{\pi}{4}\Biggr].\nonumber
\end{eqnarray}

Other slow terms emerging in the rhs of Eq.~(\ref{final}), {\em
e.g.}, the one with $\omega  \rightarrow -\omega$, do not
oscillate with magnetic field. By contrast, the function
$\text{\large F}_{\tau,T}$ {\em does} oscillate, since  the
argument of cosine in Eq.~(\ref{function1}), with
$\delta\Theta_{\mbox{\tiny B}}$ given by Eq.~(\ref{delta}), has a
{\em saddle point} at
$r_1=r_2=r_{\omega}=3C_{\omega}v_{\mbox{\tiny
F}}/4\omega_c=(3/4)C_{\omega}R_{\mbox{\tiny L}}$. In the vicinity
of the saddle point, the phase of the cosine can be presented as
\begin{eqnarray}
\label{phase} \frac{C_{\omega}\omega}{\omega_c}-\frac{\pi}{4}
-\frac{C_{\omega}\omega}{\omega_c}\Biggl[\frac{(r_1-r_{\omega})^2+
(r_2-r_{\omega})^2}{4r_{\omega}^2}\qquad\qquad\nonumber\\
+\frac{(r_1-r_{\omega})(r_2-r_{\omega})}{r_{\omega}^2}\Biggr],\qquad\qquad
\end{eqnarray}
where $C_\omega$ is defined by Eq.~(\ref{predict}). Note, that the
combination,
$\left[C_{\omega}\omega/\omega_c-\frac{\pi}{4}\right]$, in
(\ref{phase}) is nothing but the {\em phase of the
interaction-induced oscillations} Eq.~(\ref{predict}). It also
follows from Eq.~(\ref{phase}) that, when this phase is large,
the characteristic deviations, $(r_1-r_{\omega})$ and
$(r_2-r_{\omega})$ are much smaller than $r_{\omega}$. This allows
to perform the integration over these deviations in
Eq.~(\ref{function1}) {\em explicitly}. This yields
\begin{eqnarray}
\label{integrated} \text{\Large F}_{\tau,T}(\omega,\omega_c)=
\frac{1}{\left(3\cdot
2^{29/3}\pi^3\right)^{1/2}}\Biggl(\frac{\omega_c}{E_{\mbox{\tiny
F}}}\Biggr)^{5/3} \text{\Large
D}\!\left(\frac{C_{\omega}\omega}{\omega_c}\right)\times\nonumber
\\
\cos\Biggl(\frac{C_{\omega}\omega}{\omega_c}-\frac{\pi}{4}\Biggr)
A^2\left(\frac{2\pi r_{\omega}T}{v_{\mbox{\tiny
F}}}\right)\!\!A\left(\frac{4\pi r_{\omega}T}{v_{\mbox{\tiny
F}}}\right)
\exp\left[-\frac{4r_{\omega}}{v_{\mbox{\tiny F}}\tau}\right],\nonumber\\
\end{eqnarray}
where $\text{\large D}(z)=z^{-5/6}$. For high enough temperatures
$T>2\omega_c/3\pi C_{\omega}$ (but still $T\ll \omega$), the
damping factor can be replaced by the exponent, and we reproduce
the oscillating contribution  Eq.~(\ref{predict}).
\begin{figure}[t]
\centerline{\includegraphics[width=60mm,angle=0,clip]{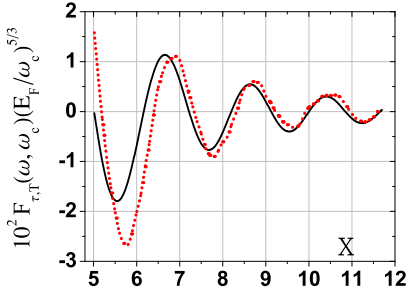}}
\caption{(Color online) Interaction-induced contribution to the ac
conductivity calculated numerically from Eq.~(\ref{function1})
(dotted line) and from asymptotic expression
Eq.~(\ref{integrated}) (full line) are plotted vs dimensionless
frequency $x=2^{1/3}\omega/(E_{\mbox{\tiny
F}}^{1/3}\omega_c^{2/3})=3(z/4)^{2/3}$. The calculations are
performed for dimensionless disorder $1/(E_{\mbox{\tiny
F}}\tau)=0.08(\omega_c/E_{\mbox{\tiny F}})^{2/3}$ and
dimensionless temperature $T/E_{\mbox{\tiny
F}}=0.06(\omega_c/E_{\mbox{\tiny F}})^{2/3}$.}
\end{figure}
The above derivation suggests that  oscillatory behavior of the
correction Eq.~(\ref{predict}) establishes only at large $z \gg
1$, which corresponds to the nodes of $\cos(z-\pi/4)$ with high
numbers. To find out where the asymptotics Eq.~(\ref{integrated})
actually applies, we have evaluated the double integral
Eq.~(\ref{function1}) numerically. The result is plotted in Fig.~3
and indicates that Eq.~(\ref{predict}) applies starting already
from the third node.

{\noindent \it Other diagrams.} Contribution
Eq.~(\ref{integrated}) to $\delta\sigma^{int}(\omega)$ is the
result of calculation of a single diagram $a$ in Fig.~ 2. Other
diagrams, involving two electron-electron scattering processes and
yielding contributions with a structure similar to
Eq.~(\ref{integrated}), are shown in Fig.~2. Diagrams $b$, $c$,
and $d$ are captured within the self-consistent Born
approximation, and correspond to certain terms in
$\delta\Sigma^{int}$, see c) in Fig.~2 (inset). Diagrams $e$-$h$
in Fig.~2 are of the same order as $a$-$d$, but they are {\em not}
contained in Eq.~(\ref{sigma}); these diagrams emerge from the
general expression Eq.~(\ref{kubo}) for $\sigma(\omega)$. Taking
all the diagrams into account leads to the modification of the
function $D(z)$ from $z^{-5/6}$ to ${\tilde
D}(z)=-32z^{-5/6}+64z^{1/6}$, where the factors $-32$ and $64$
account for the spin indices and for the number of closed fermion
loops in different diagrams. The second term, $\propto z^{1/6}$,
arises from the diagrams $b$, $e$ and $d$, $h$ in Fig.~2.
Since
the oscillations in Eq.~(\ref{predict}) develop at $z \gg 1$,
these diagrams are, actually, dominant.

 {\noindent \it Numerical estimates.} Note that, in terms of
$B$-periodicity, oscillations Eq.~(\ref{predict}) coincide with
oscillations Eq.~(\ref{standard})
 upon rescaling $\omega_c$ by $2\pi/C_{\omega}$ in
the argument of cosine, and by $2\pi/3C_{\omega}$ in the Dingle
factor. For a typical ac frequency $\hbar\omega \sim 3$K and
density $n \sim 10^{11}$cm$^{-2}$ in the experiments
\cite{zudov01,zudov03,mani02,dorozhkin03,studenikin,vitkalov}
 this shifts
the domain of oscillations Eq.~(\ref{predict}) from $B \sim 0.2$T
to $B \lesssim 10^{-2}$T. For such $B$ the observation of the
oscillations requires $T<\omega_c/6\pi C_{\omega} \sim 20$mK,
which was not the case in
Refs.~\onlinecite{zudov01,zudov03,mani02,dorozhkin03,studenikin,vitkalov}.
For observation of magneto-oscillations Eq.~(\ref{predict}) higher
densities $n\sim 5\cdot 10^{11}$cm$^{-2}$ and frequencies
$\hbar\omega \sim 15$K are needed.


\begin{thebibliography}{99}
\bibitem{footnote} For the same reason we neglect
the Zeeman splitting
\end{thebibliography}
\end{document}